\documentstyle[aps,preprint,prc]{revtex}
\def\hbw{$\hbar \omega$}

\begin{document}

\begin{center}
{\Large \bf Toward a Consistent Description of the PNC Experiments in A=18-21
Nuclei\\}
\vspace{.8cm}

{\bf Mihai Horoi$^{1,2}$ and B. Alex Brown$^{1}$\\}

\vspace{.8cm}

{\em
$^{1}$National Superconducting Cyclotron Laboratory,
East Lansing, MI 48824\\

$^{2}$Institute of Atomic Physics, Bucharest, Romania\\}

\vspace{42pt}

\end{center}

\vspace{.3cm}

\begin{abstract}

The experimental PNC results in $^{18}$F, $^{19}$F, $^{21}$Ne and the
current theoretical analysis show a discrepancy .
If one interprets the
small limit of the experimentally extracted PNC matrix element
for $^{21}$Ne as a destructive interference between
the isoscalar and the isovector contribution, then it
is difficult to understand why the isovector contribution in
$^{18}$F is so small  while the isoscalar + isovector
contribution in $^{19}$F is relatively large.
In order to understand the origin of this discrepancy
 a comparison of the calculated
PNC matrix elements was performed. It is shown that the
$^{18}$F and $^{21}$Ne matrix elements contain important contributions
from 3$\hbar \omega$ and 4$\hbar \omega$ configuration and that the
(0+1)$\hbar \omega$ calculations give distorted results.

\vspace*{.7cm}

{\bf{PACS numbers:}} 21.60.-n,21.60.Cs,27.20.+n,11.30.Er,21.10.Ky

\end{abstract}

\newpage

The investigations of low energy parity nonconservation (PNC)
phenomena in light nuclei
have as a goal to provide more reliable results for the
hadron-meson weak coupling constants. These couplings are of
importance for our understanding of the quarks behavior inside the nucleons
under the influence of the fundamental interactions. These
investigations necessitate both very delicate experiments and very
reliable nuclear structure calculations of the matrix elements
for a correct extraction of the weak nucleon-meson coupling constants.

Most of the results on the experimental and theoretical PNC studies
in light nuclei have been
presented in the
a review paper \cite{AH}. From the proposed cases during
the last 25 years in this range of nuclei, four cases have been
selected as reliable enough for experimental and theoretical
analysis. They involve parity mixed doublets (PMD)\cite{AH} in
$^{14}$N, $^{18}$F, $^{19}$F and $^{21}$Ne. Two others cases
involving PMD's in $^{16}$O \cite{KHDC} and $^{20}$F \cite{HC92}
have been  proposed recently. From the four mentioned cases, only
the $^{19}$F has been measured with a result larger then the
experimental error. All other cases have been measured with errors larger
($^{18}$F and $^{21}$Ne) or near the result ($^{14}$N). However, the
absolute value of the measured errors for $^{18}$F and $^{21}$Ne
are so small that they impose severe constraints on the different
contributions to the PNC matrix elements. These results,
compared to current
theoretical calculation have shown a discrepancy,
which has not yet been solved (see also Fig. 11 from Ref \cite{Page}).
 Namely, if one interprets the
small limit of the (experimentally) extracted PNC matrix element
($<0.029$ eV) for $^{21}$Ne as a destructive interference between
the isoscalar and the isovector contribution \cite{AH}, then it
is difficult to understand why the isovector contribution in
$^{18}$F is so small ($<0.09$ eV) and the isoscalar + isovector
contribution in $^{19}$F is relatively large (0.40 $\pm$ 0.1).

In the last years,  efforts have been made to improve
the shell-model calculations with special emphasis on the
description of the weak
observables \cite{HJ90,WBM92}. Recently, two new interactions have
been developed by Warburton and Brown \cite{WB92}, which were
designed to
accurately describe pure $\hbar \omega$ states in nuclei with A=10-22,
but methods have been developed to use them when mixed n$\hbar \omega$
excitations contribute \cite{HCBW,WBM92}. Recently, we performed a (0+1+2+3+4)
$\hbar \omega$ calculation in $^{14}$N \cite{HCBW} and we obtained a PNC
matrix element with a magnitude consistent with the experimental result
\cite{ZP89,ZA94}.
These result encouraged us to look to the effect of the higher
n$\hbar \omega$ excitations for the $^{18}$F, $^{19}$F and $^{21}$Ne cases.
The previous theoretical analysis of these cases is based on the calculations
of
Haxton \cite{AH}: (0+1+2)$\hbar \omega$ for  $^{18}$F (this
result was shown to
be comparable with the matrix element extracted almost model independent from
the first forbidden beta decay of $^{18}$Ne \cite{HPRL}) and
(0+1)$\hbar \omega$ calculations for $^{19}$F and $^{21}$Ne.

At present, is impossible to perform (0+1+2+3+4)$\hbar \omega$ calculations for
these nuclei in the full p-sd model space or any extension of it.
However, the small ZBM model space \cite{ZBM} contains the most important
PNC transition ($1p_{1/2}\ -\ 2s_{1/2}$) and moreover, includes up to
4$\hbar \omega$ excitations. We decided to analyse the results obtained in this
model space with respect to the contribution of the higher
n$\hbar \omega$ excitations and to compare, when possible, with the results
from the larger s-p-sd-fp model spaces.
We do not expect that the absolute values obtained
in this small ZBM model space to be accurate enough because we use a free space
PNC potential in a severely truncated s.p. space. One would like to use
an effective PNC potential, valid in a truncated model space,
 but this difficult problem has not been solved yet.
However, we expect that the relative values of the PNC matrix elements
 to be significant
for our analysis. Table 1 present the amplitudes of the various
n$\hbar \omega$ contributions to the many body wave functions for the parity
mixed doublets in A=18-21 nuclei. The calculations have been
performed in the ZBM
model space with the F-psd interaction \cite{PRC73}. One can see that in
all cases the 3$\hbar \omega$ contribution is significant in all negative
parity cases and 4$\hbar \omega$ contribution is relatively large for
$^{18}$F and $^{21}$Ne. Similar magnitudes can be obtained with all the
available  interactions in this model space \cite{OXBA}.

It is interesting to see if the effect of these amplitudes is reflected in
the magnitude of the PNC matrix element. Table 2 presents the
$n \hbar \omega \rightarrow (n+1) \hbar \omega$ decomposition of the PNC matrix
elements for the above mentioned nuclei. One can observe the alternation
of sign for different contributions which has been explained in Ref.
\cite{AH} in a simple Nilsson quadrupole plus pairing scheme. The message of
this behaviour is that one has to take into account an appropriate number
of n$\hbar \omega$ in order to "smooth out" this cancelation behaviour. For
instance, taking only (0+1)$\hbar \omega$ could give very distorted results
due to the missing, and opposite in sign, $1 \rightarrow 2 \hbar \omega$
contribution. This cancellation effect is particularly strong for the
$^{18}$F and $^{21}$Ne nuclei. For both those nuclei the
$2 \rightarrow 3 \hbar \omega$ and $3 \rightarrow 4 \hbar \omega$ contributions
look relatively important. Moreover, in the $^{21}$Ne case the most
important contribution is $1 \rightarrow 2 \hbar \omega$ so that the analysis
based on (0+1)$\hbar \omega$ \cite{AH}
turns out to be inappropriate for $^{21}$Ne.

One cannot fully verify this conclusions in a larger model space. However,
one can perform some tests for $^{18}$F.  A calculation for the parity mixed
doublet in this nucleus has been carried out
 using the Warburton-Brown interaction
in the first four major shells including up to 3$\hbar \omega$ excitations.
In all these calculations: (a) the spuriousity due to the center of mass
motion has been removed with the method described in Ref. \cite{GL};
(b) the effect of the short range correlations have
been taken into account as described in Ref. \cite{BPRL80};
(c) the effect of the Saxon-Woods
tail of the single particle wave functions has been checked  and
found  to be negligeable. In all our calculations we have
used the DDH best values \cite{DDH} for the weak coupling
constants. Figure 1 presents the
n$\hbar \omega$ decomposition of the wave
functions and PNC matrix element calculated in the ZBM model space with the
F-psd interaction. Figure 2 presents the same quantities calculated in the
larger model space described above. One can see that up to 20\% the
contributions looks relatively similar. These give us some confidence that
the small model space calculations contains the most important trends
necessarily for the analysis of the PNC matrix elements in light nuclei.
Certainly, one cannot rely on the absolute values given by this calculations
but they put in evidence the most important features involved
and they could indicate the way
to improve them.

As we already mentioned, the smallness of the PNC matrix element for
$^{21}$Ne has been interpreted as a cancellation between the isoscalar
and isovector contribution when only the (0+1)$\hbar \omega$ calculation
is taken into account. We have already seen that the ZBM calculations
indicate that the $1 \rightarrow 2 \hbar \omega$ contribution is the most
important one for this case. Table 3 presents the isoscalar-isovector
decomposition of the PNC matrix elements for these nuclei. One can see
that the isovector contribution is very stable in all these cases. On
the other hand, the isoscalar contribution in the $^{21}$Ne case is small
and fluctuates around a zero value. This fact, correlated with the smallness
of the pion weak coupling constant as deduced from the $^{18}$F experiment
\cite{AH,Page},
can explain the smallness of
the PNC matrix element for $^{21}$Ne as due to a very suppressed isoscalar
contribution.
A rather different conclusions has
been presented in Ref. \cite{BRAN78}, where the isoscalar contribution is
stable and the isovector contribution fluctuates. Their results do not offer
any explanation of the smallness of the PNC matrix element. A similar
conclusion to ours has been recently presented in Ref. \cite{DD93}.

One can go a step further in the analysis of all A=18-21 results using a
graphical picture, similar to  Fig. 11 from Ref. \cite{Page}.
 One can write the PNC matrix element for all these cases in terms
of the isoscalar (IS) and isovector (IV) contributions calculated with some
"standard" weak coupling constants (DDH best values in our case)
and some weighting factors, $\alpha_{IS(IV)}$ and $\beta_{IS(IV)}$

\begin{equation}
<V_{PNC}> = \alpha_{IS} \cdot \beta_{IS} <V_{PNC}^{DDH}(IS)>_{ZBM}
 + \alpha_{IV} \cdot
\beta_{IV} <V_{PNC}^{DDH}(IV)>_{ZBM} \ .
\label{eq:scalpnc}
\end{equation}

\noindent
We note that the IS matrix element is dominated by the $\rho$ exchange
term proportional to $h_{\rho}^{\circ}$ and the IV matrix element is
dominated by the $\pi$ exchange term proportional to $f_{\pi}$
\cite{AH,BPRL80}.
The $\beta$ factors take into account the renormalization effects due to
the orbitals missing in the ZBM model space.
Our assumption is that if we use the same interaction in
the ZBM model space the $\beta_{IS}$ and $\beta_{IV}$
factors will be practically the
same for all three nuclei ($^{18}$F, $^{18}$F and $^{21}$N).
The results
presented in Fig. 3 are based on the F-psd interaction \cite{PRC73}.
A value of $\beta_{IV} = 0.59$ can be obtained from the comparison
with
the $^{18}Ne$~$\rightarrow$~$^{18}F$ first forbidden beta decay result
\cite{Page}.
A value of $\beta_{IS}= 0.48$ was estimated from a comparison
with
a recent (0+1+2+3+4)$\hbar \omega$ calculation \cite{HCBW}
in $^{14}$N.
The $\alpha$ factors represent the ratio of the actual weak coupling
constants to the DDH best values \cite{DDH}.
An ($\alpha_{IS},\ \alpha_{IV}$) plot, similar to
that in Fig. 11 from Ref. \cite{Page} is presented in Fig. 3; it
shows an overlapping region for the
$^{18}$F,
$^{19}$F and $^{21}$Ne data.
The ($\alpha_{IS},\ \alpha_{IV}$) values in the
overlapping region are in the range (0.6-1.2, 0.07-0.26).
(If one extrapolates this analysis as in Ref. \cite{DD93} and assumes that
the isoscalar matrix element is compatible with zero one can conclude that the
$^{21}$Ne experiment has measured, in fact, an isovector matrix element.
This could impose smaller limits limits on the pion weak coupling
constants than the $^{18}$F experiment.)

In conclusion, we have theoretically analysed the PNC
experiments in A=18-21 nuclei using the small ZBM model space but looking
closely to the n$\hbar \omega$ contributions. We concluded that at least
for $^{18}$F and $^{21}$Ne it is necessary to estimate at least the effect
of 3 and 4 $\hbar \omega$ contributions. The present analysis suggests that
the usual interpretation of the smallness of the $^{21}$Ne matrix element
due to a cancellation between the IS and IV contribution, which was obtained
in a (0+1)$\hbar \omega$ calculation, has to be modified due to the
apparently strong $1 \rightarrow 2 \hbar \omega$ contribution. We find
that the possible interpretation is based on a very small IS part of
the PNC matrix element, due to the nuclear structure involved,
and the smallness of the IV part, due to
the small pion weak coupling constant. Our analysis show that a consistent
understanding of the PNC experiments in A=18-21 nuclei is possible
if one include the appropriate number of n$\hbar \omega$ excitations
in the nuclear structure calculations.

\vspace*{1.2cm}

The authors would like to acknowledge support from the
the Alexander von Humboldt Foundation and NSF grant 94-03666.
M.H. thanks
Soros Foundation, Bucharest, Romania for a travel
grant.



\newpage
\begin{center}
{\bf Table captions}
\end{center}

\vspace{1cm}

{\bf Table 1} \ The amplitudes of the n$\hbar \omega$ excitations to
the many body wave functions for the parity mixed doublets in
A=18-21 nuclei.\\

{\bf Table 2} \ Partial contributions to the PNC matrix elements in
A=18-21 nuclei. Units are eV. F-psd and Z-psd interactions are taken from
Ref. \cite{PRC73} and ZBMO interaction from Ref. \cite{OXBA}.
 Here and in the following calculations the DDH best values have been
used as weak coupling
constants.\\

{\bf Table 3} \ Isoscalar (IS) and isovector (IV) contributions to the
PNC matrix elements in A=18-21 nuclei.\\

\vspace{3cm}

\begin{center}
{\bf Figure captions}
\end{center}

\vspace{1.5cm}

{\bf Figure 1} \ Wave function amplitudes and partial contributions
to the isovector PNC matrix element (eV) for $^{18}$F. Calculations have
been carried out in the small ZBM model space with the F-psd interaction. \\

{\bf Figure 2} \ Same as Fig. 1 but in a larger single particle space
(first four major shells) using the Warburton-Brown interaction \cite{WB92}.
 Only
(0+1+2+3)$\hbar \omega$ excitations have been allowed to contribute.
The ? indicates uncalculated quantities\\

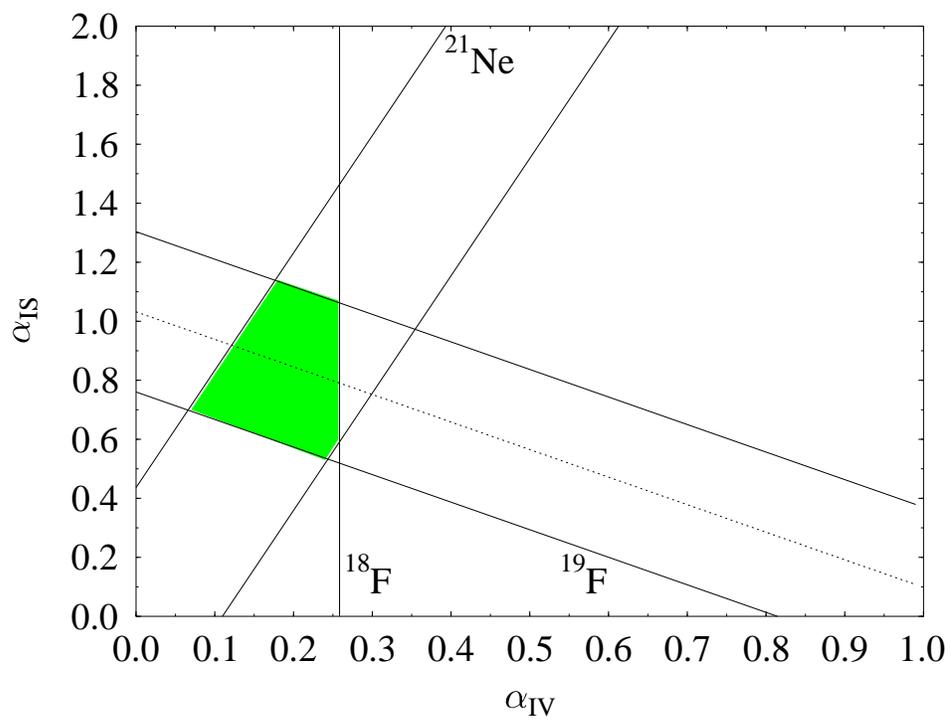
{\bf Figure 3} \ Analysis of the PNC result for $^{18}$F, $^{19}$F
and $^{21}$Ne using Eq. (\ref{eq:scalpnc}) and ZBM F-psd
calculations. Solid lines represent
the limits imposed by the experimental errors. Doted line is the experimental
result for $^{19}$F. Shadowed region indicates the consistency of the
experimental result with the present analysis.\\


\newpage
\parindent0.0em
\oddsidemargin.0cm
\evensidemargin.0cm
\topmargin0.0cm
\textheight23.5cm
\textwidth16cm
\headheight0cm
\headsep0cm

\vspace*{4cm}

\begin{center}
\begin{tabular}{ccccccc}
\hline
\hline
Nucleus & $J^{\pi}T$ & 0 \hbw & 1 \hbw & 2 \hbw & 3 \hbw & 4 \hbw \\
\hline
\hline
& $0^{+}1$ & \ \ 0.443\ \  & & \ \ 0.424\ \  & & \ \ 0.133\ \  \\
$^{18}$F & & & & & &\\
& $0^{-}0$ & & \ \ 0.739\ \  & & \ \ 0.261\ \  & \\
\hline
& $\frac{1}{2}^{+} \frac{1}{2}$ & \ \ 0.618\ \  & & \ \ 0.333\ \
 & & \ \ 0.049\ \  \\
$^{19}$F & & & & & & \\
& $\frac{1}{2}^{-} \frac{1}{2}$ & & \ \ 0.701\ \  &  & \ \ 0.299\
\  & \\
\hline
& $1^{+}1$ & \ \ 0.637\ \  & & \ \ 0.326\ \  & & \ \ 0.036\ \  \\
$^{20}$F & & & & & & \\
& $1^{-}1$ & & \ \ 0.800\ \  & & \ \ 0.200\ \  & \\
\hline
& $\frac{1}{2}^{+} \frac{1}{2}$ & \ \ 0.463\ \  & & \ \ 0.422\ \
 & & \ \ 0.115\ \  \\
$^{21}$Ne & & & & & & \\
& $\frac{1}{2}^{-} \frac{1}{2}$ & & \ \ 0.697\ \  & & \ \ 0.303\
\  & \\
\hline
\hline
\end{tabular}\\

\vspace{.4cm}
Table 1.
\end{center}

\newpage

\vspace*{3cm}

\begin{center}
\begin{tabular}{ccccccc}
\hline
\hline
Nucleus & Interaction & $\Delta$T & 0\hbw - 1\hbw & 2\hbw - 1\hbw &
 2\hbw - 3\hbw & 4\hbw - 3\hbw \\
\hline
\hline
& F-psd & 1 & 1.045 & -0.815 & 0.549 & -0.187 \\
\cline{2-7}
$^{18}$F & Z-psd & 1 & 1.119 & -0.778 & 0.462 & -0.148 \\
\cline{2-7}
& ZBMO & 1 & 1.297 & -0.669 & 0.430 & -0.118 \\
\hline
& & 0 & 0.566 & -0.097 & 0.227 & -0.073 \\
$^{19}$F & F-psd & 1 & 0.744 &  -0.212 & 0.221 & -0.032 \\
\cline{2-7}
 & & 0 & 0.633 & -0.134 & 0.184 & -0.055 \\
& Z-psd & 1 & 0.858 & -0.164 & 0.187 & -0.023 \\
\hline
& & 0 &  0.473 & -0.026 &  0.099 & -0.018  \\
$^{20}$F & F-psd & 1 & 0.806 & -0.261 & 0.195 & 0.018 \\
\cline{2-7}
&  & 0 & 0.446 & -0.086 & 0.037 & -0.007 \\
& Z-psd & 1 & 0.762 & -0.222 & 0.108 & 0.002 \\
\hline
& & 0 & 0.290 & -0.370 & 0.139 & -0.172 \\
$^{21}$Ne & F-psd & 1 & -0.164 & 0.558 &  -0.095 & 0.257 \\
\cline{2-7}
&  & 0 & 0.404 & -0.348 & 0.092 & -0.076 \\
& Z-psd & 1 & -0.246 & 0.555 & -0.053 & 0.103 \\
\hline
\hline
\end{tabular}\\

\vspace{.4cm}
Table 2.
\end{center}

\newpage

\vspace*{4cm}

\begin{center}
\begin{tabular}{ccccc}
\hline
\hline
Nucleus & Interaction & IS & IV & Total \\
\hline
\hline
& F-psd & -  & \ \ 0.592\ \  &  0.592  \\
$^{18}$F & Z-psd & - & \ \ 0.734\ \  &  0.734 \\
& ZBMO & - & \ \ 0.794\ \  &  0.794 \\
\hline
& F-psd & \ \ 0.627\ \  &  \ \ 0.722\ \  & 1.349  \\
$^{19}$F & Z-psd & \ \ 0.629\ \  & \ \ 0.858\ \ & 1.487 \\
\hline
& F-psd & \ \ 0.518\ \  &  \ \ 0.757\ \  &  1.262  \\
$^{20}$F & Z-psd & \ \ 0.389\ \ & \ \ 0.650\ \ & 1.032 \\
\hline
& F-psd & \ \ -0.113\ \  &  \ \ 0.556\ \  & 0.442  \\
$^{21}$Ne & Z-psd & \ \ 0.071\ \  & \ \ 0.359\ \ & 0.430  \\
& ZBMO & \ \ -0.010\ \  &  \ \ 0.339\ \  & 0.329 \\
\hline
\hline
\end{tabular}\\

\vspace{.4cm}
Table 3.
\end{center}

\newpage
\parindent0.0em
\oddsidemargin.0cm
\evensidemargin.0cm
\topmargin0.0cm
\textheight23.5cm
\textwidth16cm
\headheight0cm
\headsep0cm

\vspace*{3.5cm}

\begin{center}

\begin{minipage}[t]{12cm}
\unitlength1.cm

\begin{picture}(10.,10.)(0,0)
\thicklines
\put(0.5,8.5){\makebox(0,0)[bl]{$\vert ^{18}$F(0$^{+}1)_{1}\ >$\ \
 =\ \ 0.666$\vert 0\ \hbar \
\omega >$\ +\ 0.651$\vert 2\ \hbar \ \omega >$
\ +\ 0.365$\vert 4\ \hbar \ \omega >$}}

\put(0.3,5.){\makebox(0,0)[bl]{$<V_{PNC}^{\Delta T=1}>_{DDH}$
=\ \ \ \ \ 1.045\ \ -\ \ \ 0.815\ \ \ \ \ \ +\ \ \ \ \ 0.549\ \ -\ \ 0.187}}

\put(0.5,1.5){\makebox(0,0)[bl]{$\vert ^{18}$F(0$^{-}0)_{1}\ >$\ \
=\ \ \ 0.860$\vert 1\ \hbar \
\omega >$\ \ \ \ \ \ \ \ \ \ +\ \ \ \
\ \ \ \ \ \ \ 0.511$\vert 3\ \hbar \ \omega >$}}

\put(4.9,8.2){\line(0,-1){2.8}}
\put(7.7,8.2){\line(-1,-3){.94}}
\put(8.3,8.2){\line(1,-3){.94}}
\put(10.9,8.2){\line(0,-1){2.8}}

\put(4.9,4.7){\vector(0,-1){2.8}}
\put(6.46,4.7){\vector(-1,-3){.94}}
\put(9.45,4.7){\vector(1,-3){.94}}
\put(10.9,4.7){\vector(0,-1){2.8}}

\end{picture}
\begin{center}
Figure 1.
\end{center}
\end{minipage}

\end{center}

\newpage

\vspace*{3.5cm}

\begin{center}

\begin{minipage}[t]{12cm}
\unitlength1.cm

\begin{picture}(10.,10.)(0,0)
\thicklines
\put(0.5,8.5){\makebox(0,0)[bl]{$\vert ^{18}$F(0$^{+}1)_{1}\ >$\ \
 =\ \ 0.744$\vert 0\ \hbar \
\omega >$\ +\ 0.668$\vert 2\ \hbar \ \omega >$
\ +\ \ ?\ $\vert 4\ \hbar \ \omega >$}}

\put(0.3,5.){\makebox(0,0)[bl]{$<V_{PNC}^{\Delta T=1}>_{DDH}$
=\ \ \ \ \ 0.921\ \ -\ \ \ 0.502\ \ \ \ \ \ +\ \ \ \ \ 0.533\ \ -\ \ \
\ ? }}

\put(0.5,1.5){\makebox(0,0)[bl]{$\vert ^{18}$F(0$^{-}0)_{1}\ >$\ \
=\ \ \ 0.743$\vert 1\ \hbar \
\omega >$\ \ \ \ \ \ \ \ \ \ +\ \ \ \
\ \ \ \ \ \ \ 0.669$\vert 3\ \hbar \ \omega >$}}

\put(4.9,8.2){\line(0,-1){2.8}}
\put(7.7,8.2){\line(-1,-3){.94}}
\put(8.3,8.2){\line(1,-3){.94}}
\put(10.9,8.2){\line(0,-1){2.8}}

\put(4.9,4.7){\vector(0,-1){2.8}}
\put(6.46,4.7){\vector(-1,-3){.94}}
\put(9.45,4.7){\vector(1,-3){.94}}
\put(10.9,4.7){\vector(0,-1){2.8}}

\end{picture}

\begin{center}
Figure 2.
\end{center}
\end{minipage}

\end{center}


\begin{thebibliography}{99}

\bibitem{AH}
E.G. Adelberger and W.C. Haxton, Ann. Rev. Nucl. Part. Sci.
${\bf 35}$, 501 (1985).

\bibitem{KHDC}
N. Kniest, M. Horoi, O. Dumitrescu and G. Clausnitzer,
Phys. Rev. ${\bf C 44}$, 491 (1991).

\bibitem{HC92}
M. Horoi and G. Clausnitzer, Phys. Rev. {\bf C48}, R522(1993).


\bibitem{Page}
S. A. Page et al., Phys. Rev. {\bf C35}, 1119(1987).

\bibitem{ZP89}
V.J. Zeps et al., A.I.P. Conf. Proc. {\bf 176}, 1098 (1989).




\bibitem{ZA94}
V. J. Zeps, E. G. Adelberger, A. Garc\'{\i}a, C.A. Gossett,
H. E. Swanson, W.Haeberli, P.A. Quin and J. Sromicki, to be published


\bibitem{HCBW}
M. Horoi, G. Clausnitzer, B.A. Brown and E.K. Warburton,
Phys. Rev. {\bf C} - in press.

\bibitem{WB92}
E.K. Warburton and B.A. Brown, Phys. Rev. {\bf C46}, 923 (1992).

\bibitem{HJ90}
W.C. Haxton and C. Johnson, Phys. Rev. Lett. {\bf 65}, 1325 (1990).

\bibitem{WBM92}
E.K. Warburton, B.A. Brown and D.J. Millener, Phys. Lett. {\bf
293B}, 7 (1992).

\bibitem{DDH}
B. Desplanques, J. F. Donoghue and B. R. Holstein, Ann. Phys. (N. Y.)
{\bf 124}, 449(1980).


\bibitem{HPRL}
W.C. Haxton, Phys. Rev. Lett. {\bf 46}, 698(1981).

\bibitem{ZBM}
A.P. Zucker, B. Buck and J.B. McGrory, Phys. Rev. Let. {\bf 21}, 39(1968).


\bibitem{PRC73}
J.B. McGrory and B.H. Wildenthal, Phys. Rev. {\bf C7}, 974(1973).


\bibitem{BPRL80}
B.A. Brown, W.A. Richter and S. Godwin, Phys. Rev. Lett. {\bf 45}, 1681(1980).


\bibitem{OXBA}
B.A. Brown et al., MSUNSCL Report, ${\bf 524}$  (1988).

\bibitem{GL}
D.H. Gloeckner and R.D. Lawson, Phys. Lett. {\bf 53B}, 313 (1974).\\
P.R. Rath, A. Faessler, H. Muther and A. Watt, J. Phys.
{\bf G 16}, 245 (1990).



\bibitem{BRAN78}
R.A. Brandenburg et al., Phys. Rev. Lett. {\bf 41}, 618(1978).

\bibitem{DD93}
B. Desplanques and O. Dumitrescu, Nucl. Phys. {\bf A565}, 818(1993).


\end{thebibliography}
\end{document}